\documentclass[pra,aps,twocolumn,nobalancelastpage,superscriptaddress,colorinlistoftodos]{revtex4-2}

\usepackage[utf8]{inputenc}
\usepackage[T1]{fontenc}
\usepackage[english]{babel}
\usepackage{csquotes}
\usepackage{comment}

\usepackage{amsmath}
\usepackage{amsfonts}
\usepackage{relsize}
\usepackage{bm}

\usepackage[margin=2cm]{geometry}

\usepackage{siunitx}
\DeclareSIUnit{\bits}{bits}

\usepackage{graphicx}
\graphicspath{{./figures/}}
\usepackage{pgf}
\usepackage{float}

\everymath=\expandafter{\the\everymath\displaystyle}
\usepackage[caption=false]{subfig}

\usepackage{array}
\usepackage{multirow}
\usepackage{makecell}

\usepackage[hidelinks]{hyperref}
\usepackage{natbib}
\usepackage{glossaries-extra}
\setabbreviationstyle[acronym]{long-short}

\newacronym{qkd}{QKD}{Quantum Key Distribution}
\newacronym{pqc}{PQC}{Post-Quantum Cryptography}
\newacronym{otp}{OTP}{One-Time Pad}
\newacronym{cow}{COW}{Coherent One Way}
\newacronym{cwdm}{CWDM}{Coarse Wavelength Division Multiplexing}
\newacronym{nat}{NAT}{Network Address Translation}
\newacronym{vpn}{VPN}{Virtual Private Network}
\newacronym{qms}{QMS}{Quantum Management System}
\newacronym{kms}{KMS}{Key Management System}
\newacronym{voa}{VOA}{Variable Optical Attenuator}
\newacronym{kem}{KEM}{Key Encapsulation Mechanism}
\newacronym{qber}{QBER}{Quantum Bit Error Rate}
\newacronym{qrng}{QRNG}{Quantum Random Number Generator}

\usepackage[textsize=tiny]{todonotes}
\makeatletter
\define@key{todonotes}{Yoann}[]{%
    \setkeys{todonotes}{author=Yoann,color=red}}%

\makeatother

\usepackage{mdframed}

\mdtheorem{mdprotocol}{Protocol}

\date{}

\usepackage{layouts}
\begin{document}
\title{Quantum Key Distribution with Efficient Post-Quantum Cryptography-Secured Trusted Node on a Quantum Network}

\author{Yoann Piétri}
\affiliation{Sorbonne Université, CNRS, LIP6, F-75005 Paris, France}

\author{Pierre-Enguerrand Verdier}
\affiliation{Orange Innovation, F-92326 Châtillon, France}

\author{Baptiste Lacour}
\affiliation{Orange Innovation, F-92326 Châtillon, France}

\author{Maxime Gautier}
\affiliation{Orange Innovation, F-92326 Châtillon, France}

\author{Heming Huang}
\affiliation{Telecom Paris, Institut Polytechnique de Paris, F-91120 Palaiseau, France}

\author{Thomas Camus}
\affiliation{ID Quantique SA, CH-1227 Genève, Switzerland}

\author{Jean-Sébastien Pegon}
\affiliation{ID Quantique SA, CH-1227 Genève, Switzerland}

\author{Martin Zuber}
\affiliation{CryptoNext Security, F-75005 Paris, France}

\author{Jean-Charles Faugère}
\affiliation{CryptoNext Security, F-75005 Paris, France}

\author{Matteo Schiavon}
\affiliation{Sorbonne Université, CNRS, LIP6, F-75005 Paris, France}

\author{Amine Rhouni}
\affiliation{Sorbonne Université, CNRS, LIP6, F-75005 Paris, France}

\author{Yves Jaou{\"e}n}
\affiliation{Telecom Paris, Institut Polytechnique de Paris, F-91120 Palaiseau, France}

\author{Nicolas Fabre}
\affiliation{Telecom Paris, Institut Polytechnique de Paris, F-91120 Palaiseau, France}

\author{Romain All\'eaume}
\affiliation{Telecom Paris, Institut Polytechnique de Paris, F-91120 Palaiseau, France}

\author{Thomas Rivera}
\affiliation{Orange Innovation, F-92326 Châtillon, France}

\author{Eleni Diamanti}
\affiliation{Sorbonne Université, CNRS, LIP6, F-75005 Paris, France}

\date{\today}

\begin{abstract}
    Quantum Key Distribution (QKD) enables two distant users to exchange a secret key with information-theoretic security, based on the fundamental laws of quantum physics. While it is arguably the most mature application of quantum cryptography, it has inherent limitations in the achievable distance and the scalability to large-scale infrastructures. While the applicability of QKD can be readily increased with the use of intermediary trusted nodes, this adds additional privacy requirements on third parties. In this work, we present an efficient scheme leveraging a trusted node with lower privacy requirements thanks to the use of post-quantum cryptographic techniques, and implement it on a deployed fiber optic quantum communication network in the Paris area.
\end{abstract}

\maketitle

\section{Introduction}

The significant expected progress in algorithmic techniques and computing power in the next years, including using powerful quantum processors, has brought to the forefront the need for developing quantum-safe cryptographic solutions. Such solutions may advantageously combine techniques leveraging mathematical algorithms believed to be robust against quantum attacks, namely \gls{pqc}, and techniques exploiting quantum resources, in particular \gls{qkd}. The latter is an ensemble of methods and protocols that allows two distant users, usually called Alice and Bob, equipped with an untrusted public quantum channel and a public authenticated classical channel, to exchange a random string of bits with information-theoretic security. This security, based on the fundamental laws of quantum physics, ensures that this bitstring can later be used as a symmetric cryptographic key.

The field of \gls{qkd} has seen remarkable progress in the last year~\cite{AdvancesPirandola2020}. However, the limited achievable distance and the difficulty in scaling to large networks remain important practical challenges in \gls{qkd} implementations. The limit in range is due to the fundamental law allowing \gls{qkd} in the first place: an unknown quantum state cannot be cloned and since photon transmission decays exponentially in fibers, the achievable distance is limited in theory by well-established bounds~\cite{Pirandola2017}. Furthermore, \gls{qkd} requires point-to-point communication for the exchange of the quantum states, which is rather unpractical in large networks. To overcome this issue, it is possible to use intermediary nodes with optical switches, a technique known as physical bypass, but doing so is detrimental to the key rate and does not extend the range of communication.

A practical solution to these limitations is to break the communication link into several sublinks and perform \gls{qkd} on each sublink. The final key is then relayed by the intermediary nodes using the QKD keys. This solution has already been implemented in several quantum communication networks, but lowers the security of the key exchange: first because the trusted node is an additional location for the malicious adversary, Eve, to physically attack, and second because of the full trust that has to be accorded to the intermediary node, which will directly be in possession of the final key.

Hybrid solutions, combining \gls{pqc} and \gls{qkd} techniques, can be used to mitigate the practical security challenges encountered in \gls{qkd} implementations. Several proposals targeting, for instance, the authentication step or the aforementioned trusted node security issue, have been made~\cite{Walenta2015,Muckle, Muckle-plus,battarbee2024quantumsafehybridkeyexchanges,HybridQkdAmpGeitz2023,zeng2024practicalhybridpqcqkdprotocols}. In this work, we also address the trusted node security issue; in particular, by combining the standard trusted node relay protocol with a post-quantum key encapsulation mechanism and AES encryption, we show that we are able to lower the trust requirement on the intermediary node. Our approach is similar to the one presented in~\cite{HybridQkdAmpGeitz2023}, however our protocol features a better efficiency in terms of key bit usage, saturating the final secret key rate. We also demonstrate its practical relevance by implementing it in a deployed optical fiber quantum communication network.

The paper is structured as follows: in section~\ref{sec:trusted-node} we present the standard trusted node protocol and the modified version we will implement. Then, in section~\ref{sec:quantum-network} we present the Paris Quantum Network, along with the \gls{qkd} setup, before describing the results of the experiment in section~\ref{sec:results} and drawing some conclusions in section~\ref{sec:conclusion}.

\section{Trusted node Protocol\label{sec:trusted-node}}

Let us start by presenting the considered setup and the underlying assumptions. For this, we first introduce the following \textit{notations}: if $k$ is a key binary register of $n$ bits, we denote by $|k| = n$ the size of the register, and for $0 \leq i \leq n$, by $\lfloor k \rfloor_i$ the truncated key up to the $i$th term, which is a key register of $i$ bits. Additionally, we use the symbol $\oplus$ that refers to the bitwise modulo-2 addition between two registers.

\subsection{Setup}

Alice and Bob are two trusted users who want to exchange a key, while Charlie is an intermediary node. Eve is a malicious adversary wanting to learn the content of the key.

Alice and Bob are both linked to Charlie with a public quantum channel and a public authenticated classical channel. Alice and Bob are also linked with a public authenticated classical channel. We will denote by $QC(\text{Node 1}, \text{Node 2})$ (resp. $CC(\text{Node 1}, \text{Node 2})$) the public quantum channel (resp. authenticated classical channel) linking Node 1 and Node 2.

Alice, Bob and Charlie have the required hardware to run the \gls{qkd} protocols, and in particular we suppose that Alice and Charlie can perform \gls{qkd} with $QC(A,C)$ and $CC(A,C)$ and that Charlie and Bob can perform \gls{qkd} with $QC(B,C)$ and $CC(B,C)$ (possibly with different \gls{qkd} protocols and/or implementations). This also means that we make the standard assumptions in \gls{qkd} implementations: Alice, Bob and Charlie have access to secure locations, trusted quantum and classical hardware, and true random number generators. Additionally, we make the assumption that the parties are bounded by the laws of quantum physics.

Another standard assumption in \gls{qkd} is that Alice and Bob are behaving honestly, and follow the protocol instructions. As for Charlie, we want to introduce here more nuances by considering three possible honesty levels: Charlie could be \textit{honest}, by blindly following the protocol instructions without leaking information or remembering what he sees, he could be \textit{honest-but-curious} (or \textit{semi-honest}), where he follows the protocol instructions but attempts to learn as much information as possible from the received messages and he could be \textit{malicious}, where he can deviate from the protocol with no constraints. In the last case, we can say that Charlie is controlled by the adversary Eve.

\subsection{Standard trusted node protocol}

The standard trusted node protocol goes as follows:\\

\begin{mdprotocol}[QKD with a trusted node]
    Alice, Bob and Charlie have access to a \texttt{QKD} subroutine.

    \begin{enumerate}
        \item Alice and Charlie perform \texttt{QKD} using $QC(A,C)$ and $CC(A,C)$. They both end up with a key $k_{AC}$ of length $l_{AC}$.
        \item Bob and Charlie perform \texttt{QKD} using $QC(B,C)$ and $CC(B,C)$. They both end up with a key $k_{BC}$ of length $l_{BC}$.
        \item Bob communicates the value of $l_{BC}$ to Alice over the classical channel $CC(A,B)$.
        \item If $l_{AC} = 0$ or $l_{BC} = 0$, Alice makes the protocol abort, otherwise she computes $l=\min(l_{AC}, l_{BC})$. Alice communicates $l$ to Bob and Charlie over the classical channels $CC(A,B)$ and $CC(A,C)$.
        \item Alice generates the random key $k_{AB}$ of length $l$ and computes $m_1 = k_{AB}\oplus \lfloor k_{AC}\rfloor_l$. Alice sends $m_1$ to Charlie over the classical channel $CC(A,C)$.\label{item:otp1}
        \item Charlie recovers the key $k_{AB}$ by $k_{AB} = m_1\oplus \lfloor k_{AC}\rfloor_l$ and computes $m_2 = k_{AB}\oplus \lfloor k_{BC}\rfloor_l$. Charlie sends $m_2$ to Bob over the classical channel $CC(B,C)$.\label{item:otp2}
        \item Bob recovers the key $k_{AB}$ by $k_{AB} = m_2\oplus \lfloor k_{BC}\rfloor_l$.\label{item:otp3}
    \end{enumerate}
    Alice and Bob end up with the key $k_{AB}$ of length $l = \min(l_{AC}, l_{BC})$.
\end{mdprotocol}

Since the combination of \gls{qkd} with \gls{otp}, corresponding to steps \ref{item:otp1}, \ref{item:otp2} and \ref{item:otp3} in the protocol, achieves perfect secrecy, the messages $m_1$ and $m_2$ cannot be deciphered to recover $k_{AB}$ by the adversary Eve, in the cases where Charlie is honest or semi-honest.

However, in this protocol, Charlie directly holds the key $k_{AB}$ meaning that he could decipher all messages exchanged between Alice and Bob that were encrypted using this key, by simply monitoring the classical channels. This means that the standard trusted node protocol allows no protection against an honest-but-curious Charlie.

As mentioned earlier, another downside of trusted node \gls{qkd} is that it introduces an additional location that can be attacked by the adversary. While this is not an issue when considering the standard assumptions of \gls{qkd}, since we are considering that no information comes out of the secure locations except for the quantum and classical channels, it could be an issue in some practical \gls{qkd} scenarios. Indeed, Alice and Bob may be reticent to trust the security of Charlie's location over which they might have no control.

\subsection{Modified trusted node protocol}

Let us now present the modified trusted node protocol. For this, we need to introduce the notion of a \glsxtrfull{kem}. This is a protocol that allows to exchange a cryptographic key (usually to be used in symmetric encryption protocols) over a public channel, using asymmetric encryption. We say that it is a \gls{pqc}-\gls{kem} if the asymmetric mechanisms rely on post-quantum cryptographic methods. 

The modified protocol goes as follows:

\begin{mdprotocol}[\gls{qkd} with \gls{pqc}-secured trusted node\label{prot:modified-trusted-node}]
    Alice and Bob have access to a \texttt{PQC-KEM} subroutine. Alice, Bob and Charlie have access to a \texttt{QKD} subroutine.
    \begin{enumerate}
        \item Alice and Bob use \texttt{PQC-KEM} to exchange the symmetric key $k_{AES}$ over the public $CC(A,B)$.
        \item Alice and Charlie perform \texttt{QKD} over their quantum and classical channels $QC(A,C)$ and $CC(A,C)$. They both end up with a key $k_{AC}$ of length $l_{AC}$.
        \item Bob and Charlie perform \texttt{QKD} over their quantum and classical channels $QC(B,C)$ and $CC(B,C)$. They both end up with a key $k_{BC}$ of length $l_{BC}$.
        \item Bob communicates the value of $l_{BC}$ to Alice over the classical channel $CC(A,B)$.
        \item If $l_{AC} = 0$ or $l_{BC} = 0$, Alice makes the protocol abort, otherwise she computes $l=\min(l_{AC}, l_{BC})$. Alice communicates $l$ to Bob and Charlie over the classical channels $CC(A,B)$ and $CC(A,C)$.
        \item Alice generates the random key $k_{AB}$ of length $l$ and encrypts it using the encryption function $k^{enc}_{AB} = \texttt{ENC}_{\text{AES}}(k_{AES}, k_{AB})$ and computes $m_1 = k^{enc}_{AB}\oplus \lfloor k_{AC}\rfloor_l$. Alice sends $m_1$ to Charlie over the classical channel $CC(A,C)$.
        \item Charlie computes $k^{enc}_{AB} = m_1\oplus \lfloor k_{AC}\rfloor_l$ and $m_2 = k^{enc}_{AB}\oplus\lfloor k_{BC}\rfloor_l$. Charlie sends $m_2$ to Bob over the classical $CC(B,C)$.
        \item Bob computes $k^{enc}_{AB} = m_2\oplus\lfloor k_{BC}\rfloor_l$ and decrypts the key $k_{AB} = \texttt{DEC}_{\text{AES}}(k_{AES}, k^{enc}_{AB})$.
    \end{enumerate}
    Alice and Bob end up with the key $k_{AB}$ of length $l = \min(l_{AC}, l_{BC})$.
\end{mdprotocol}

\begin{figure*}
    \centering
    \includegraphics[width=0.7\textwidth]{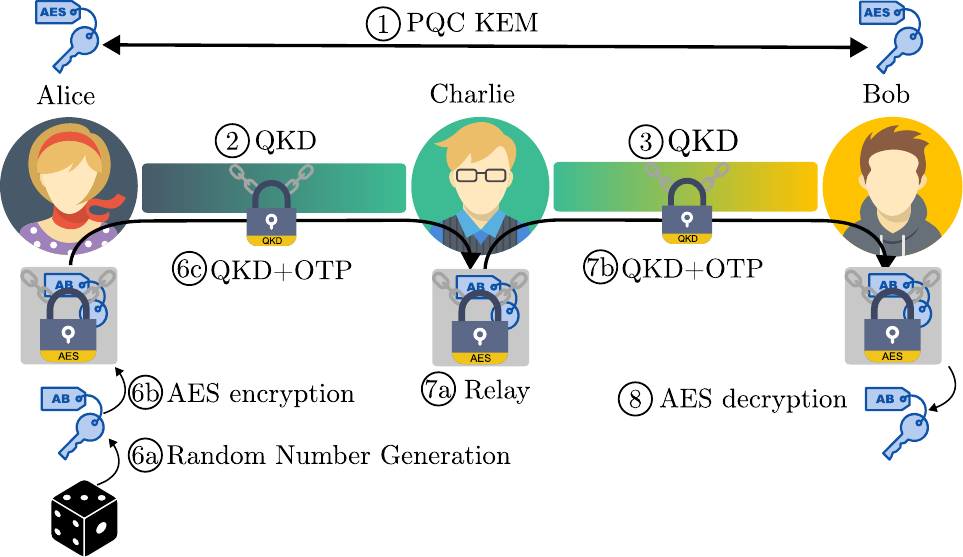}
    \caption{Schematic representation of the protocol. Numbers match the steps indicated in Protocol~\ref{prot:modified-trusted-node}. Classical communications over classical channels are represented with black arrows, while QKD operation is represented on the gradient rectangle. For simplicity, some classical communications were omitted from the scheme, in particular the classical communication for QKD and steps 4 and 5 of the protocol.}
    \label{fig:protocol}
\end{figure*}

The protocol is represented in a schematic way in Fig.~\ref{fig:protocol}.

This modified protocol increases the difficulty of an attack for an honest-but-curious Charlie. Indeed, he would have to recover the key $k_{AES}$ to get the final key. This provides computational security against an honest-but-curious trusted node.

This method does not help against physical attacks against Charlie's location. Indeed, an unbounded adversary could recover the $k_{AES}$ encryption by breaking the \gls{pqc}-\gls{kem} and then by recovering $k_{AB}^{enc}$ at the location of Charlie. Simply obtaining $k_{AES}$ is however not sufficient to get the final key without attacking Charlie's location.

\subsection{Efficiency analysis}

Here we compare the efficiency of the presented protocol with respect to the one in~\cite{HybridQkdAmpGeitz2023}, where the final key $k_{AB}$ is directly encrypted and decrypted using the \gls{pqc}-\gls{kem} algorithm Crystals-Kyber. In more details, the protocol goes as follows: 1. Alice (called WFD01 in~\cite{HybridQkdAmpGeitz2023}) generates a random key using an ID Quantique \gls{qrng}; 2. The random key is encrypted with Kyber using the public key of Bob (WFD03)~\footnote{There is an additional step of authentication using the Falcon signature scheme that we do not consider here.}; 3. The \gls{pqc} encrypted key is encrypted again using \gls{otp} and the \gls{qkd} key shared with Charlie, and the ciphertext is send to Charlie (WFD02) over the public channel; 4. Using the key shared with Alice, Charlie decrypts the ciphertext (getting the \gls{pqc}-encrypted version) and re-encrypts it using \gls{otp} (this time with the \gls{qkd} key shared with Bob) before sending it to Bob; 5. Bob decrypts the ciphertext using the key shared with Charlie and the private \gls{pqc} key to recover the final secret.

Let $l$ be the target number of bits in the final key $|k_{AB}| = l$. The question is how many bits from the keys exchanged with QKD are necessary for the modified trusted node protocol.

The operation of Kyber is defined by a security parameter~\cite{kyber-website}. This parameter relates to the security performance and also impacts the size of the ciphertext in the \gls{kem}. Indeed, for an input key size of $\SI{256}{\bits}$, the ciphertext length, that we will denote $l_{ct}$, is $\SI{6144}{\bits}$ for Kyber-512, $\SI{8704}{\bits}$ for Kyber-768 and $\SI{12544}{\bits}$ for Kyber-1024~\cite{kyber-website}.

In~\cite{HybridQkdAmpGeitz2023}, the output ciphertext of the KEM is directly encrypted using \gls{otp}. For simplicity, let's suppose that the final key length $l$ is a multiple of $256$ (in practice, the ID Quantique Cerberis system that we will use for our demonstration stores the key in $\SI{256}{\bit}$ blocks), so $l = 256p$ for $p\in\mathbb{N}$. This means that the final key will be encrypted in $p$ blocks of $l_{ct}\,\si{\bits}$, each one of them using \gls{otp} and the keys distributed using \gls{qkd}. Hence, we can compute the ratio of the final key length with the number of bits used for \gls{otp}:
\begin{equation}
    \eta = \frac{l}{p\times l_{ct}} = \frac{256}{l_{ct}}.
\end{equation}

Here we compute the ratio with respect to the key consumption on one \gls{qkd} link. In this way, the value of $\eta$ will be directly used to derive the final key rate as $r_{\rm final} = \eta\min(r_{AC}, r_{BC})$ where $r_{\rm final}, r_{AC}, r_{BC}$ are respectively the secret key rate of the final key exchange, \gls{qkd} exchange between Alice and Charlie, and \gls{qkd} exchange between Bob and Charlie (this assumes that both \gls{qkd} links are running simultaneously). Note that, when considering a key relay with perfect secrecy with respect to the outside world, as is the case here, the maximal final secret key rate is bounded by $\min(r_{AC}, r_{BC})$ since as many bits of \gls{qkd} key as in the final key are required for perfect secrecy. In this sense, $r_{\rm final}\leq \min(r_{AC}, r_{BC})$ and $\eta \leq 1$. The value of $\eta$ for the different Kyber parameters are given in Table~\ref{tab:efficiency}.

In comparison, when we first exchange a random $\SI{256}{\bits}$ key using the Kyber \gls{kem} on a public channel and use this key for AES-256, the following happens: AES-256 uses a block size of $\SI{128}{\bits}$, and each ciphertext of a $\SI{128}{\bit}$-block is also of length $\SI{128}{\bits}$. This means, that our final key of length $l = 256p$ is encrypted into $2p$ blocks of length $\SI{128}{\bits}$ giving the efficiency
\begin{equation}
    \eta = \frac{l}{2p\times 128} = 1.
\end{equation}

Changing the parameter of the Kyber protocol will induce a longer key which will result in a higher number of bits exchanged on the public channels (in the first step), which is not a bottleneck.

\bgroup
\def\arraystretch{1.5}
\begin{table}
    \centering
    \begin{tabular}{|c||c|c|c||c|c|c|}
    \hline
         Protocol & \multicolumn{3}{c||}{Protocol in \cite{HybridQkdAmpGeitz2023}} & \multicolumn{3}{c|}{Our protocol}\\
         \hline
         Kyber & 512 & 768 & 1024 & 512 & 768 & 1024 \\
         \hline
         $\eta$ & $\SI{4.17}{\percent}$ & $\SI{2.94}{\percent}$ & $\SI{2.04}{\percent}$ & \SI{100}{\percent} & \SI{100}{\percent} & \SI{100}{\percent}\\
         \hline
    \end{tabular}
    \caption{Efficiency depending on the protocol and Kyber parameter.}
    \label{tab:efficiency}
\end{table}
\egroup

If the key has a length that is not a multiple of $128$ or $256$, there is an added inefficiency due to the required padding, but this tends to $0$ as the key length grows, since the number of padded bits is always less than $128$ or $256$.

We believe that the security of the protocol is not  impacted in a significant way by our modifications compared with the protocol in~\cite{HybridQkdAmpGeitz2023}. In particular, in Tab.~\ref{tab:security} we explicit the protocol(s) that need to be broken by Charlie (or Eve) to gain access to the final secret key.

\bgroup
\def\arraystretch{1.5}
\begin{table*}
    \centering
    \begin{tabular}{|c|c|c||c|}
    \hline
     Protocol & Charlie (Honest-But-Curious) & Eve & $\eta$ \\
     \hline
     Usual Trusted Node & Nothing & OTP+QKD & 100\%\\
     \hline
     Protocol in~\cite{HybridQkdAmpGeitz2023} & PQC-KEM & OTP+QKD and PQC-KEM & 2-4\%\\
     \hline
     Our protocol & PQC-KEM or AES & OTP+QKD and (PQC-KEM or AES) & 100\% \\
     \hline
\end{tabular}
    \caption{Protocol(s) to break in order to gain access to the final key.}
    \label{tab:security}
\end{table*}
\egroup

Note that in all three cases, Eve has to break the OTP+QKD exchange (assuming all other locations secure). Additionally in~\cite{HybridQkdAmpGeitz2023} or our protocol, she also has to break either the PQC-KEM or AES (this second one only in our version). Note that in comparison with~\cite{HybridQkdAmpGeitz2023}, in our protocol another option is given to both Charlie and Eve with respect to breaking the PQC-KEM, which is to break AES. However, it is believed that breaking a symmetric protocol such as AES will still remain harder than breaking an asymmetric one, even with a quantum computer.

\section{The Paris Quantum Network\label{sec:quantum-network}}

Next, we describe the Paris Quantum Network, where we performed the experimental demonstration of our protocol, the characteristics of the links and the devices that were used for the \gls{qkd} exchanges.

\subsection{Physical infrastructure}

\begin{figure*}
    \subfloat[On-scale map of the quantum network. The actual fiber links do not correspond to the straight depicted line. The nodes and links of interest here are in red. Nodes and links in blue were not part of the implementation of this protocol. An interactive version of this map can be found at \url{https://u.osmfr.org/m/1051066}.]{\includegraphics[width=0.5\textwidth]{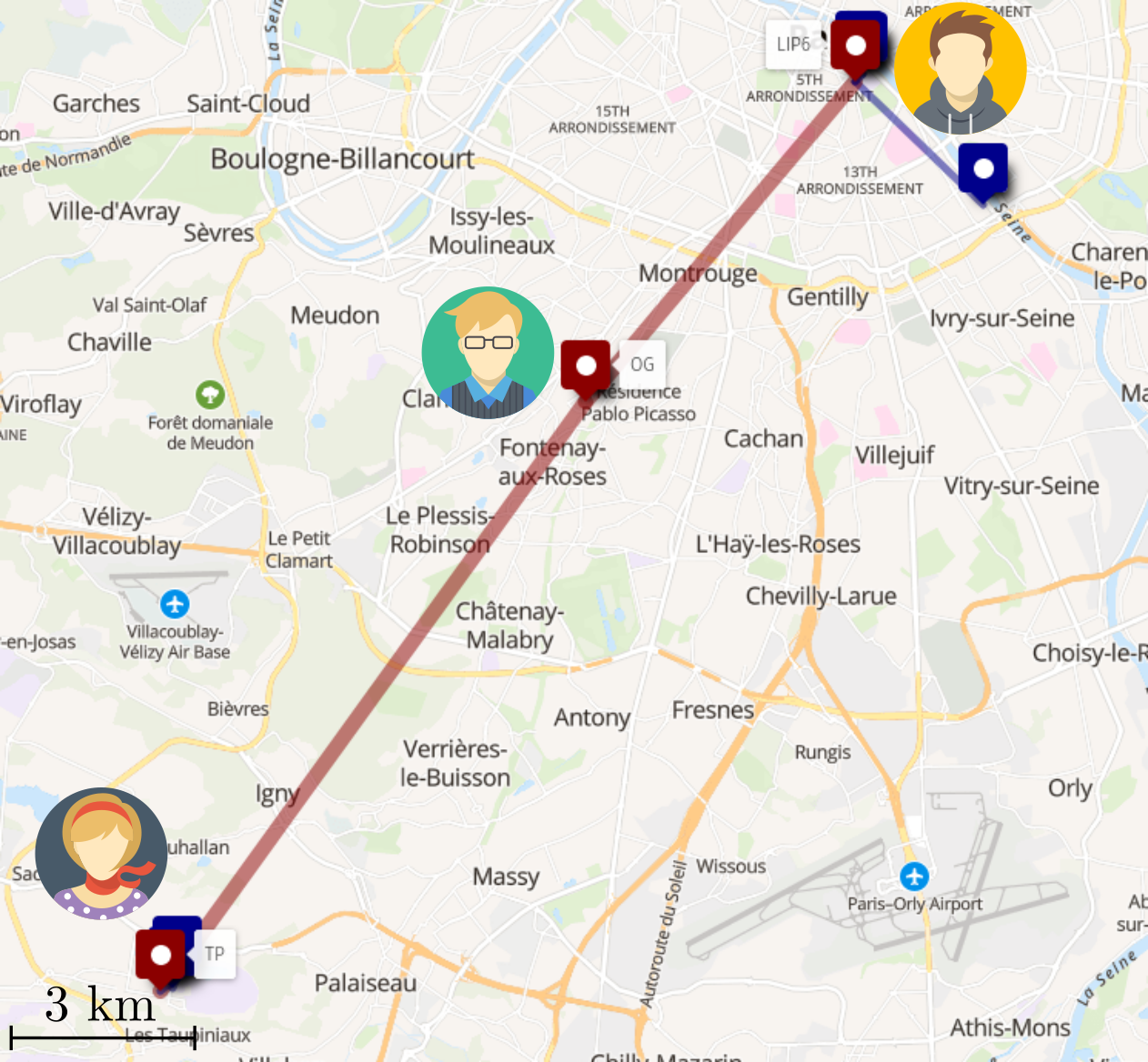}}%
    \hfill
    \subfloat[Graph-like representation of the quantum network. The nodes of interest are in red. This figure is not to scale. The length of the fibers have been indicated.\label{subfig:network}]{\includegraphics[width=0.32\textwidth]{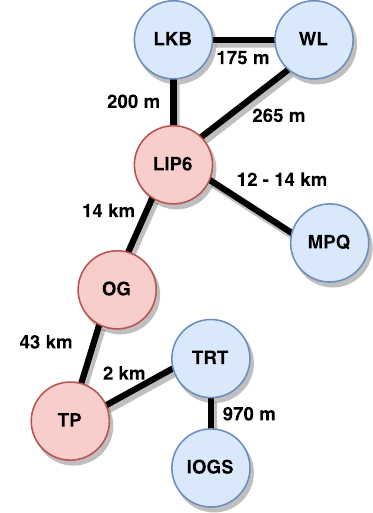}}
    \caption{The Paris Quantum Network.}
    \label{fig:quantum-network}
\end{figure*}

The physical infrastructure of the Parisian Quantum Network is currently composed of 8 nodes, located in the Paris Region, as shown in Fig.~\ref{subfig:network}.

The three nodes of interest here are LIP6 (LIP6, Sorbonne Université, in the 5th district of Paris), OG (Orange Innovation in Châtillon) and TP (Télécom Paris, Institut Polytechnique de Paris in Palaiseau). For completeness, the other connections are described in more detail in appendix~\ref{appendix:full-network}. The connections are done using dark fibers dedicated to quantum communication applications, which are standard SMF-28 fibers. The network was assembled by splicing existing segments that were once used for classical communications. Between LIP6 and OG, two fibers of length $\SI{14}{\kilo\meter}$ (average losses of $\SI{3.8}{\deci\bel}$) are available and between OG and TP, two fibers of length $\SI{43}{\kilo\meter}$ (average losses of $\SI{10.4}{\deci\bel}$) are available.

\subsection{QKD systems}

The \gls{qkd} systems are the commercial devices Cerberis XGR from ID Quantique~\cite{CerberisXGR}. They are performing the \gls{cow} protocol~\cite{ContinuousHighStucki2009} using time-bin qubits.

Each \gls{qkd} system is composed of two nodes of standard size 1U, one containing the transmitter (Alice) and the other one containing the receiver (Bob). The two nodes need to be connected with an optical fiber serving as the quantum channel. Moreover, they also need to be connected by one or two optical fiber(s) allowing for full duplex communication between them, for synchronization. The fiber(s) should be about the same length as the quantum channel fiber. Finally the two nodes also need to be addressable with direct IPv4 links, and by the central management software, hosted in Châtillon.

We implement the full-duplex synchronization channel by using a single optical fiber and bi-directional modules (Skylane Optics SBHEDB22L32D and SBHEUB22L32D) that use the \gls{cwdm} technology to have one channel (\gls{cwdm} high) in $[\lambda_T + \SI{1.5}{\nano\meter}, \lambda_T + \SI{6.5}{\nano\meter}]$ and the second channel (\gls{cwdm} low) in $[\lambda_T - \SI{6.5}{\nano\meter}, \lambda_T - \SI{1.5}{\nano\meter}]$ where $\lambda_T$ is the central operating wavelength. In our case, we choose a wavelength close to the one used in the quantum channel $\lambda_T = \SI{1550}{\nano\meter}$.

To perform the trusted node scheme, two \gls{qkd} systems are needed (and hence 4 nodes) and we will refer to them as Pair $1$ and Pair $2$ with the nodes Alice $1$, Bob $1$, Alice $2$ and Bob $2$. The first pair operates with an attenuation up to $\SI{18}{\deci\bel}$ and the second pair with an attenuation up to $\SI{12}{\deci\bel}$. Due to the asymmetry of our links, we choose to deploy Pair $1$ on the TP-OG link ($\SI{43}{\kilo\meter}$, $\SI{10.4}{\deci\bel}$) since with the additional connectors it would be close to or surpass the $\SI{12}{\deci\bel}$ limit. Pair $2$ is deployed on the OG-LIP6 link ($\SI{14}{\kilo\meter}$, $\SI{3.8}{\deci\bel}$).

\subsection{Classical network}

Since direct IPv4 addressing is required to operate the Cerberix XGR systems, the solution of using a \gls{vpn} was chosen. Hence a \gls{vpn} was established between the three remote locations over the internet, using the Wireguard software~\cite{Wireguard} . Each node was equipped with a router to operate the \gls{vpn} and communicate with the local equipment. Here, we stress that the \gls{vpn} was established for routing purposes only and that the security of the \gls{qkd} exchanges and the overall key exchange does not rely on the inherent security provided by the \gls{vpn}.

Additionally, the router in OG was also the central management node of the \gls{qkd} nodes used to deploy the configuration and collect statistics. This was done using the \gls{qms} solution of ID Quantique.

\subsection{Encryptors}

Keys were retrieved by interacting with the \gls{kms} of each node using the standardised communication protocol ETSI-QKD-014~\cite{ETSI-QKD-104}. In practice, this was done using a specifically modified version of the IP9001 Mistral encryptors from Thales~\cite{Mistral}. The encryptors were used to encrypt the data of a 4K video streaming service~\footnote{The encryptors were using the final key to perform AES256 encryption of the streaming data.}.

\section{Results\label{sec:results}}

For the implementation of the \gls{pqc}-\gls{kem}, the Crystals-Kyber Key Encapsulation Mechanism was chosen~\cite{8406610}, and implemented by CryptoNext Security~\cite{cqsl}. Crystals-Kyber is based on the Learning-With-Errors (LWE) problem, and was submitted, along with many others, to the NIST Post-Quantum Cryptography Standardization process. It is the only one selected to be standardized for key establishment~\cite{nist_selected_2024} (as the ML-KEM algorithm in the FIPS 203 standard~\cite{fips203}). 

\begin{figure*}
    \centering
    \includegraphics{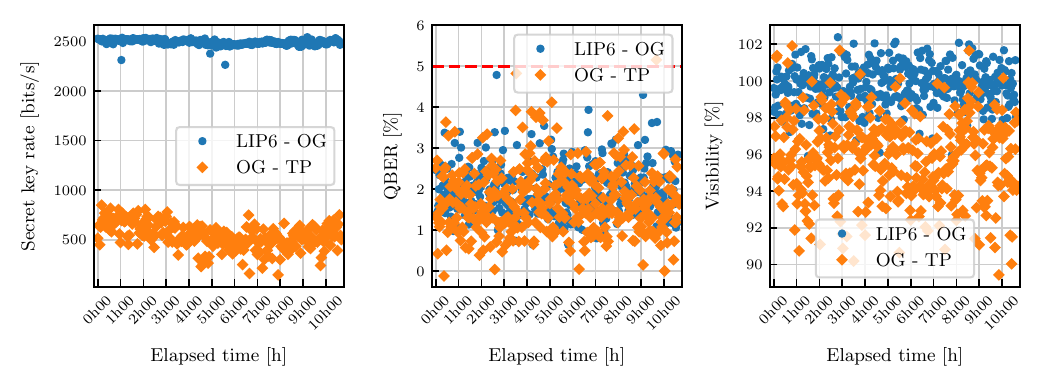}
    \caption{Secret key rate, QBER and visibility as a function of time for the two QKD links.}
    \label{fig:results}
\end{figure*}

The \gls{qkd} systems were operating in the Paris quantum network during several weeks, and were used for the trusted node experiment during one week. In Fig.~\ref{fig:results}, we show the performance of the \gls{qkd} exchanges for the last $\SI{11}{\hour}$ of the experiment, by plotting the secret key rate, \gls{qber} and visibility given by the \gls{qkd} systems.

The average \gls{qber}s were respectively $\SI{1.93}{\percent}\pm \SI{0.57}{\percent}$ (OG-LIP6) and $\SI{1.72}{\percent} \pm \SI{0.68}{\percent}$ (OG-TP) with average visibilities of $0.998\pm 0.012$ and $0.959\pm 0.024$ respectively over the 11 hours. The relatively low visibility on the OG-TP link could be due to a misalignement in fiber on the Cerberis module interface but does not change the results on the performance of the trusted node protocol. The average key rates were respectively $\SI{2493}{\bit\per\second}$ (standard deviation $\SI{28}{\bit\per\second}$) and $\SI{612}{\bit\per\second}$ (standard deviation $\SI{139}{\bit\per\second}$). Since the heavy operations can be parallelised with the \gls{qkd} key exchanges, there is no overhead and the final key rate is given by $r_{\rm final} = \min(r_{LIP6-OG}, r_{OG-TP}) = r_{OG-TP}$. This yields an overall LIP6 - TP final key rate of $\SI{612}{\bit\per\second}$ on average.

The keys exchanged between Alice and Bob were then retrieved using the specifically modified Thales Mistral encryptors to encrypt a streaming service with 4K videos.

While we implemented the protocol on key exchange with one intermediary node, the protocol can be extended to more trusted nodes, maintaining a single \gls{kem} round and AES encryption and decryption, with all the intermediary nodes forwarding the AES ciphertext.

\section{Conclusion\label{sec:conclusion}}

In this work we presented the implementation of an efficient \gls{pqc}-secured trusted node protocol. While providing computational security against an honest-but-curious intermediary node, we maximise the efficiency of the protocol by saturating the bound for information-theoretic security to exterior adversaries.

The performance of the overall protocol could be improved, for instance by switching to the ID Quantique Clavis XGR using decoy-state BB84, other DV-QKD systems or Continuous-Variable (CV) \gls{qkd}, which could yield higher key rates on the \gls{qkd} links. This work however already readily demonstrates an important use case where bringing together quantum and post-quantum cryptographic techniques provides increased practical security in real-world configurations.

\section*{Acknowledgments}

The authors acknowledge financial support from the Île-de-France Region under the ParisRegionQCI project, the European Union’s Horizon Europe research and innovation program under the Grant Agreement No 101114043 (QSNP), and the PEPR integrated project
QCommTestbed, ANR-22-PETQ-0011, which is part of Plan France 2030

Figures were created using illustrations from the Flat Profile Avatar collection (CC BY Ceria Studio), from the Travel Duotone Icons collection (PD Anita Csillag), the Sports And Games Icooon Mono Vectors collection (PD Icooon Mono) and from the Security 11 collection (CC0). Map was created using the umap tool with tiles from jawgmaps and map data from OpenStreetMap.

\bibliography{bibliography, others}

\appendix

\section{Full description of the Paris Quantum Network\label{appendix:full-network}}

The Paris Quantum Network is currently composed of 11 nodes, where connections endpoints are available, corresponding to locations of academic and industrial partners. Since some nodes are administrated by the same partner, we considered in the main text the 8 main nodes correspoding to individual partners. The 8 partners are: Laboratoire Matériaux et Phénomènes Quantiques in Université Paris Cité, in the 13th district of Paris (Node MPQ), Laboratoire LIP6 in Sorbonne Université, in the 5th district of Paris (Node LIP6) with 2 endpoints (LIP6 and LIP6 2), Laboratoire Kastler-Brossel in Sorbonne Université in the 5th district of Paris (Node LKB) with two endpoints (LKB and LKB 2), Welinq company in the 5th district of Paris (Node WL) with two endpoints (WL and WL2), Orange Innovation group of the French network operator Orange in Châtillon (Node OG), Laboratoire Traitement et Communication de l'Information in Télécom Paris, in Palaiseau (Node TP), Thales Research and Technology division of the company Thales in Palaiseau (Node TRT) and Laboratoire Charles Fabry in Institut d'Optique Graduate School in Palaiseau (Node IOGS).

\begin{figure}
    \centering
    \includegraphics[scale=0.67]{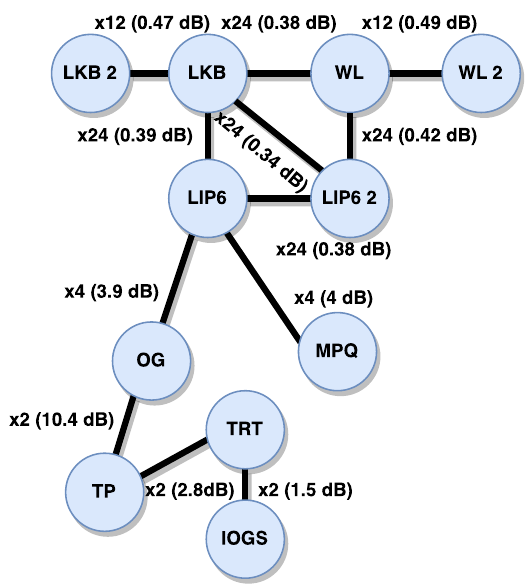}
    \caption{Complete description of the Quantum Communication backbone in the Parisian area. The labels on the edges indicate the number of the available fibers and their average losses.}
    \label{fig:network-all}
\end{figure}

For completeness, we include a full map of the current network in Fig.~\ref{fig:network-all}, including the number of available fibers and the average losses on those fibers.

This network is used to benchmark quantum technologies, in particular Quantum Key Distribution, including Discrete-Variable and Continuous-Variable, interoperability between systems, coexistence with classical communication and protocols built on top of \gls{qkd}. Moreover the network will be used for entanglement distribution and deployment of links integrating quantum memories.
\end{document}